# Machine Learning Algorithms In User Authentication Schemes

Laura Pryor
*Department of Computer Science*
*University of Wisconsin – Eau Claire*
Eau Claire, WI, USA
pryorlk8701@uwec.edu

Dr. Rushit Dave
*Department of Computer Science*
*University of Wisconsin – Eau Claire*
Eau Claire, WI, USA
daver@uwec.edu

Dr. Jim Seliya
*Department of Computer Science*
*University of Wisconsin – Eau Claire*
Eau Claire, WI, USA
seliyana@uwec.edu

Dr. Evelyn Sowells Boone
*Department of Computer Science*
*North Carolina A&T State University*
Greensboro, NC, USA
sowells@ncat.edu

*Abstract*—In the past two decades, the number of mobile products being created by companies has grown exponentially. However, although these devices are constantly being upgraded with the newest features, the security measures used to protect these devices has stayed relatively the same over the past two decades. The vast difference in growth patterns between devices and their security is opening up the risk for more and more devices to easily become infiltrated by nefarious users. Working off of previous work in the field, this study looks at the different Machine Learning algorithms used in user authentication schemes involving touch dynamics and device movement. This study aims to give a comprehensive overview of the current uses of different machine learning algorithms that are frequently used in user authentication schemas involving touch dynamics and device movement. The benefits, limitations, and suggestions for future work will be thoroughly discussed throughout this paper.

I. INTRODUCTION

The technological advancements of the past two decades regarding mobile devices has led society to develop a heightened dependency on the devices in which such personal information is now being stored. However, the security trusted to keep this information safe has stayed stagnant since the beginning of this technological boom. Most devices today still rely on static security methods to authenticate users. Static methods include entering a password, PIN number, or physiological biometrics such as a palm print [1]. While these methods are easy and reliable for the genuine user, their ease has opened the door for more and more attacks by nefarious users leaving users personal and private information vulnerable. This imminent threat on devices has led researchers to investigate possible improvements to security. One possible way of improvement is by using behavioral biometrics as an added layer of protection for devices.
Instead of focusing on passwords correctly being inputted, behavioral biometrics uses the behaviors of the user while they are interacting with their device. Behavioral biometrics has been implemented in multiple different schemas and in multiple different devices or systems. For example, [2] looks at behavioral biometrics in healthcare, and papers [3-5] look at behavioral biometrics in mobile devices. In order to use behavioral biometrics, the user authentication schema needs to include a machine learning algorithm for classification. Machine learning algorithms have been deemed very beneficial for use in cybersecurity as seen in [6] and can be very useful in behavioral biometric based models. The machine learning algorithms used in this schema often are crucial for the model to be able to perform well. That being said, while there is plenty of research available comparing the different behavioral biometrics, there is little literature done specifically on the machine learning algorithms these models depend on. Along with that, the literature that is being done on the algorithms themselves often look at multiple models using multiple different biometrics. However, some algorithms work differently with different biometrics, thus this study will focus on a common pairing of biometrics: touch dynamics and phone movement. This study aims to create a comprehensive review on the commonly used machine learning algorithms in touch dynamic based user authentication methods.

II. BACKGROUND

This literature review for machine learning algorithms used in touch dynamic and phone movement-based authentication models stems from the findings of our past work on a separate survey paper. In this paper [7], different behavioral biometric based user authentication models were investigated, and it was concluded that the biometric that yielded the most consistent and best results was touch dynamics. Also found in this study was that using multiple biometrics in a model also improved the model's accuracy. While that paper also included an analysis on the machine learning algorithms being used, many of the papers were outdated for the topic and the analysis included algorithms from models that did not include touch dynamics, so it felt necessary to complete a second literature review this time looking specifically at the machine learning algorithms for touch dynamics and phone movement-based models. Four machine learning algorithms are highlighted in this paper, those of which being Support Vector Machine (SVM), Random Forest (RF), K-Nearest Neighbor (K-NN), and Naïve Bayes (NB). SVM is a supervised machine learning algorithm that is widely used in touch dynamic and phone movement-based authentication schemas, for example in papers [8-10]. How this algorithm works is that based on the number of features used in the model, an n-dimensional space is created, and, in that space, multiple hyperplanes are found. In this hyperplane all of the data points are classified, in this case the datapoints would be classified as genuine or imposter. The goal of the algorithm is to choose the hyperplane that has the maximum distance between the two classified datapoints groups, from there all future data points are then classified based to how close the fall to the genuine or imposter "sides" of the hyperplane. RF is again another widely used machine learning algorithm in all types of authentication schemes. RF can be seen being tested in papers [11-16]. The basic premise of RF is that multiple decision trees are built and each time a new datapoint is run through the algorithm it goes through each decision tree and the results of each tree are then averaged and is then outputted as the final result. Each of the decision trees in the RF are slightly different, which adds an extra level of analysis which is missing from using just one decision tree. KNN is an algorithm that is seeing performing consistently throughout different schema [17]. Similar to SVM, KNN deals with the distance between classified points and the newly added datapoints. The K in KNN stands for a predetermined set of neighbors that are analyzed when a new datapoint is added. So, for example, if the K was 5, when a new datapoint is entered the five closest datapoints to the new point are chosen and then whatever classification is found in the majority of those five points is then "voted" as the classification for the new point. The final algorithm surveyed in this paper is NB is regarded as a very simple and fast algorithm, which can be seen in [18]. NB, unlike the other three algorithms, is a probabilistic approach. This algorithm uses the Bayes Theorem, which is used for calculating conditional probabilities. In this algorithm each feature is deemed to be both independent and equal. Therefore, in the terms of our model, the algorithm is looking for the probability that the user is genuine based on other probabilities that are known to the model based on the previous data that it has been trained on.

III. LITERATURE REVIEW

A. Support Vector Machine

In [19], researchers test SVM, along with other Machine Learning algorithms in a model that looks to authenticate users based on their touch dynamics when using social media applications. Along with SVM, the J48 Decision Tree (J48), Radial Basis Function Network (RBFN), Back Propagation Neural Network (BPNN), and NB algorithms were also tested in this schema. SVM was found to outperform RF when looking at results for testing involving long-term and continuous versus single-use authentication. The researchers used two different scenarios to test both questions, the first scenario being that data was collected during the subject's entire time on their device, and the second scenario collected data only when the user was on social media applications. When looking at short-term authentication, SVM had easily the best average error rate (AER) of 6.02% for scenario 1 and 3.07% for scenario 2. The next best algorithm was the BPNN whose results can be seen on Fig. 1. Overall, SVM had approximately a 4% lower error rate for each scenario. This paper also showed promising results for using SVM to achieve long term accuracies in user authentication schemes. After testing a group of participants on scenario 2 over a two-week span, the researchers found that the AER only increased slightly to 3.68%. The authors also noted that they noticed trending in their data that would suggest that users would be able to achieve more stable behavior changes after two weeks, which would lower that average error rate.

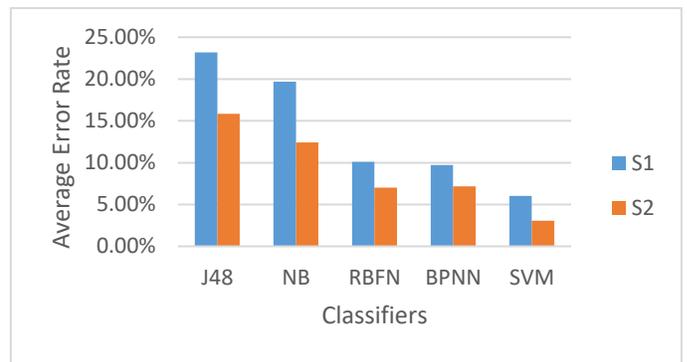

Fig. 1. Average Error Rates of the different Classifiers [19]

Researchers in [20] look to analyze the effects that behavior-based profiling systems such as WiFi and application usage have on authentication schema when combined with touch dynamics. In this paper only SVM was tested, however the results proved too strong to be ignored. The authors looked at the accuracy (ACC) for two scenarios: one time authentication and active authentication. The findings, as shown in Fig. 2, concluded that when adding behavior-based profiling systems to the behavioral biometric based schema the ACC of the models increased by an average of 17% for both one time authentication and active authentication. For one time authentication the model achieved an ACC of 82.2% when using the combined method of behavior-based profiling systems and for active authentication that same ACC jumped to 97.1%. For comparison, when testing the single behavioral biometrics for one time authentication, the highest ACC came

TABLE I. PERFORMANCE OF ALGORITHMS FOR AUTHENCATION [21]

| Success Metrics | Left Swipe | | | | | Right Swipe | | | | |
|---|---|---|---|---|---|---|---|---|---|---|
| | *SVM* | *RF (Number of Trees Below)* | | | | *SVM* | *RF (Number of Trees Below)* | | | |
| | *N/A* | *10* | *20* | *30* | *40* | *N/A* | *10* | *20* | *30* | *40* |
| Accuracy | 79.88% | 87.69% | 87.50% | 87.69% | 87.89% | 57.81% | 86.32% | 85.93% | 87.69% | 87.50% |
| FAR | 15.84% | 1.33% | 1.56% | 0.89% | 0.89% | 17.63% | 1.56% | 2.00% | 0.66% | 0.44% |
| FRR | 50.00% | 89.06% | 89.06% | 92.18% | 90.62% | 62.50% | 98.43% | 98.43% | 93.75% | 96.87% |

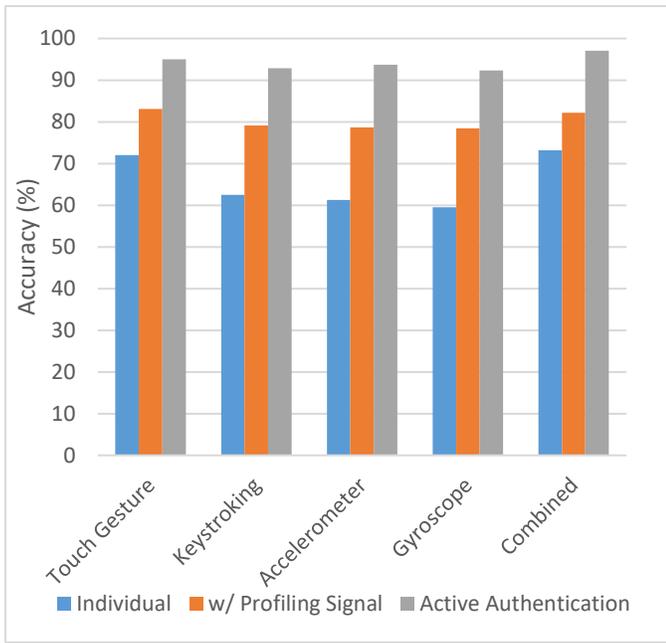

Fig. 2. Performance of model using individual biometric and the biometric paired with the profiling signals both for single use and active authentication [20]

from touch dynamics with 72.0%. These results show that SVM is an effective algorithm to use for authentication, both for single use and implicit use. The researchers also suggested that the low ACCs that came from the individual biometrics can be explained by limited numbers of data samples during some sessions. This could indicate that SVM may not be a good algorithm to use for smaller samples. Another thing to note is that this data was collected over a two-month period, providing further evidence to support that SVM performs well for long-term authentication.

In [21], SVM and RF were two algorithms used in the researcher's schema that used swipe data from touchscreens to authenticate users. In this study, the authors were experimenting with the use of the capacitive frames that are created when a user swipes on their device, instead of using touch data from the device's sensors. While SVM did not always perform the best in all scenarios, the researchers found it to be the most consistent of the two algorithms and only performed slightly less effectively than the RF algorithm in some tests. As seen in Table 1, while using SVM the model was able to produce maximum authentication accuracy of 79.88% which indicates that SVM can perform well when dealing with models that use capacitive data and not sensor data. However, using SVM also produced a false rejection rate (FRR) of 50%, and while this is much lower than the 90.62% FRR associated with RF, the high FRR still gives cause for concern for the effectiveness of the model. Researchers hypothesized that the high FRRs may be happening due to their dataset only consisting of 160 samples from 8 users, and also the training and testing data including more imposter samples than there were genuine samples. All of these reasons can lead to the model falsely rejecting the user, and it also once again indicates

Funding provided by University of Wisconsin -Eau Claire's Office of Research and Supported Projects

that, as seen in [20], SVM may not perform well with models that will be using small amounts of data to make their decision.

### B. Random Forest

Researchers Zhang et. al [22] used RF to experiment with the efficacy of their proposed continuous authentication schema. Along with RF the other algorithms being tested are SVM, NB, and decision tree (DT). In the proposed schema, four implicit swipe gestures are being used to classify the user: a vertical scroll up and down and a horizontal scroll left and right. The researchers are looking to see if using just vertical or horizontal swipe gestures omit the best results or if a combination of the two can achieve the highest accuracies. Three different models were tested, their results can be seen in Fig. 3, and RF was the best performing algorithm for each of the three models. RF achieved the highest accuracies in each of the models getting 96.7%, 96.36%, and 95.03% for the vertical swipe model, horizontal swipe model, and the combined model respectively. The results of this study indicated that the RF algorithm can be used in implicit user authentication to obtain high accuracies. These results also showed that either horizontal or vertical swipe gestures can be used in touch dynamic based schema, however the two gestures should not be combined as it decreases the model's accuracy. It was also noted by the authors that the horizontal or vertical swipe gestures could be combined with other touch dynamics to create a possibly better performing model, as the swipe gestures would not negatively affect the accuracy of the model.

In [23], the authors based their proposed model on the pattern lock that is often seen as an authentication measure for many mobile devices. In this model, a combination of the pattern lock and a password is used to create a simple game that is used for authentication purposes. In order for the user to be granted access to their device they not only need to put in the right password, but they need to put in the password in the same pattern as when they originally created it. This simple game combines the two most common forms of authentication, passwords and patterns, and puts them together, putting multiple letters and symbols included and not included in the password onto the device's screen and the user must connect the letters and symbols of the password using the correct swiping pattern. RF along with SVM and XGBoost were used in this model for testing, and the researchers found that RF was able to achieve the best results getting a False Acceptance Rate of only 1.40% and a False Rejection Rate of 2.08%. The researchers also tested the security of the model and found that even when the imposter users were given the simplest password tested, 50% of the users were not able to gain access to the device. The rest of the security results can be seen in Table 2. This implies that the extra biometric layer was able add additional protection without lengthening the process it takes to authenticate the user. Also noted was that as the password strength increased, the number of imposters able to infiltrate the device, even with previous knowledge of the password, decreased. One final item examined by the researchers was the minimum number of strokes needed for their model to authenticate the user still effectively. It was found that this model only needed four strokes, or five letters or symbols, to be able to accurately authenticate the user. Coupling that information with the low FAR and FRR of the RF classifier, it can be implied that the RF algorithm can be used in models that use a small sample size.

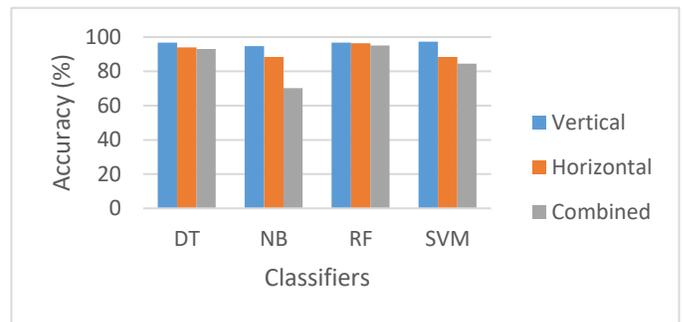

Fig. 3. Performance of the classifiers for each of the vertical, horizontal, & combined models [22]

TABLE II. SECURITY PERFORMANCE OF THE MODEL [23]

| Password Level | Security Results | | | |
|---|---|---|---|---|
| | Without Instructions | | With Instructions | |
| | Guessed | Not Guessed | Guessed | Not Guessed |
| Level 1 | 30% | 70% | 48% | 52% |
| Level 2 | 6% | 94% | 15% | 85% |
| Level 3 | 9% | 91% | 18% | 82% |
| Level 4 | 18% | 82% | 21% | 79% |

In [24], researchers Wang et. al look at multiple aspects of mobile user authentication schemes and examine how they affect the model's accuracy. One of the aspects looked at is the machine learning algorithm being used as the classifier for the schema. Six algorithms were tested in this study, those being logistic regression (LR), NB, KNN, DT, RF, and SVM. The six different models were tested 21 times with the number of training samples increasing each time, as the authors also wanted to see what the minimum number of training samples is needed to create an accurate model. RF once again performed the best of the six algorithms achieving an accuracy of 97%. It should be noted however, that all algorithms performed well, with the worst-performing algorithm, SVM, still achieving an accuracy of 81%. In this experiment it was also found that after the training sample amount reached 4, the accuracies of the six models started to level off, thus proving that RF does not need a large training sample to still be able to perform well. The researchers then continued their experiment using the RF classifier to look at feature importance. The results concluded that spatial features such as the start and stop positions of a touch gesture hold the strongest effect on the schemas efficiency, and temporal and movement direction features were found to have the weakest effect on the schema. While the authors noted that these findings are not reasoning to completely avoid using temporal and movement direction features, however it does suggest that more researchers should focus on using spatial features in their model, which ultimately increases usability as less features are needed to be collected to create an accurate decision.

## C. Other Algorithms

In [25] the machine learning algorithm KNN was highlighted. In this study the authors use KNN to test their schema that combines both the touch and movement data of the device. The results of this study are compared to four previous studies that worked with a similar schema but used different machine learning algorithms for the classifier. The researchers first testing their model by calculating the EERs of 30 unlocking gestures, 20 of which were common gestures and the other 10 were user generated. As seen in Table 3, for the first 20 common gestures, the average EER of the model was 5.24% and the average EER of the final 10 generated gestures was 4.23%. This ends up being an overall average EER of 4.90%. The familiarity of gestures was then tested as researchers were looking to see if the accuracy of the model improves when the user becomes familiar with the gestures. The EER of a user who created their own gesture was compared to the average EER of the other 40 users who were unfamiliar to the gesture and the results indicated that familiarity with the gesture increases the performance of the model. For example, after training the gesture for a week, one user achieved and EER of 2.52%, while the average of the other 40 users was 4.08%. These results could imply that K-NN could be used in long-term authentication schemas, as the EER continued to stay low for the authentic user after a week of use. The researchers also tested the minimum training samples needed for an effective authentication scheme, concluding that a sample size of 15 was the smallest amount that produced similar results to the results from models with more training samples. These results indicate the models that use K-NN do not need to use large training sets, which helps with data collection. Finally, the long-term accuracy of the model was tested, and it was found that as time increases, the EER of the model also increases slightly. While the accuracy of the model does decrease over time, it is only slightly, indicating that models that use K-NN could potentially be used for long-term authentication use, as long as the researchers create a schema that can be implicitly adapted and updated over time. The overall EER of this model was compared to the EERs of four other studies using algorithms such as SVM, RF, dynamic time warping, and MHD. This study achieved the significantly lower results than the other models. Some examples of the EERs of the other models include the model using SVM producing an EER of 9.67% and the model using RF getting an EER of 13.09%. Comparing results of the previous works to the 4.90% EER of the current model indicates that this model and the K-NN classifier can achieve significantly lower results than similar models using other major machine learning classifiers.

TABLE III. AVERAGE EERs OF THE 30 TESTED GESTURES [25]

| Gesture Type (Average) | Equal Error Rate |
|---|---|
| Common (1-20) | 5.24% |
| User-generated (21-30) | 4.23% |
| Overall | 4.90% |

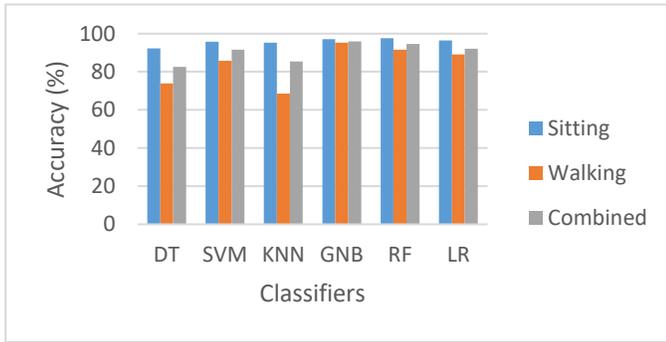

Fig. 4. Average accuracies of the classifiers [26]

Authors of [26] proposed a schema similar to others involving collecting touch dynamics while users unlocked a pattern lock, however one big difference with this schema is that instead of testing users with multiple passwords, each user had the same exact pattern. Thus, the accuracy of the proposed model comes solely from the behavioral-biometric layer rather than the biometric layer and knowledge layer combined. Six machine learning algorithms were tested in this model, those being DT, SVM, KNN, gaussian naïve bayes (GNB), RF, and LR. In the researchers first test to find the best performing classifier, the six algorithms were put against each other with the model's accuracy while the user was sitting, walking, and sitting and walking combined being collected. GNB was deemed the best performing algorithm achieving 97.19%, 95.25%, and 95.97% accuracies for sitting, standing, and combined respectively, which can be seeing in Fig 4. The next best algorithm was RF, which also produced similar results, however GNB was chosen due to the optimal feature length for GNB being shorter than RF. Thus, GNB was not only a well performing algorithm, but a cheap algorithm for the researchers to implement. The next experiment conducted by the researchers was to find the optimal number of imposter samples needed for each of the three postures (sitting, walking, and combined), since the amount of imposter users could lead to an imbalance of the FRR and FAR. After testing multiple combinations of imposter user sample sizes, the results concluded that the best imposter counts for each posture were 13 for sitting, 23 for walking, and 8 for the combined postures. The researchers also noted that this data was all collected in one day, so the model would most likely become less accurate as time increased. The results of this study indicate that the GNB algorithm can be a well-performing classifier for touch gesture-based schema.

## IV. DISCUSSION & ANALYSIS

In this survey, there were many different machine learning algorithms used in touch dynamic based user authentication schemes introduced and discussed. These algorithms were compared to each other in table [table number]. While many of the algorithms performed well in their respective models, two algorithms excelled among the rest. As seen in the sample of studies discussed, random forest and support vector machine are two heavily used and strong performing algorithms. While many of the other algorithms, such as KNN and NB, were also used quite frequently, they were often outperformed by SVM and RF which is seen throughout the study.

### A. Support Vector Machine

SVM's consistent success in mobile user authentication models could be due to the algorithm's ability to capture more complex relationships in the data than many other machine learning algorithms can. However, one downside to SVM found during our research was that SVM often requires larger training samples to be able to perform well, which in turn would increase the number of samples needed from users thus making the model very computationally costly.

### B. Random Forest

Unlike SVM, RF was found to perform well with smaller datasets, which can lead to a low-cost model. Also, RF is often considered a strong algorithm for researchers due to the algorithm's usability and forgiveness. RF is an easy algorithm for researchers to understand and implement making it very usable and the algorithm is very forgiving since an error in one tree will not negatively affect the entire model, unlike some other algorithms.

## V. LIMITATIONS & FUTURE WORK

While all of the schemes studied in this survey produced very promising results, there are still some limitations that can be affecting these accuracies. Many studies noted using small amounts of participants, which can cause issues with the consistency and validity of their models. Having invalid or inconsistent models can produce false accuracies, both negatively and positively, therefore the results shared by most of these papers could be inaccurate. Along with that, most of the models did not distinguish single touch and multi touch gestures to be different from one another, which can lead to more false positives. Finally, an issue plaguing this study and not the individual papers is that not every model had the same success metrics being used, while it is still pretty easy to evaluate the effectiveness of each of the models, it is hard to compare models that have different metrics for success. Future work should look into increasing participant sizes and creating models that can distinguish different gestures as solving these issues could increase the accuracy of the results being published and would make these model's that much closer to being ready for real-world application.

TABLE IV. COMPARISON OF THE RESULTS OF THE PAPERS SURVEYED

| Papers | ML Algorithm Summaries | | |
|---|---|---|---|
| | *Classifiers Tested* | *Best Classifier* | *Best Results* |
| [8] | SVM | SVM | 97.40% ACC |
| [9] | SVM, KNN, NB | SVM | 1-2% EER |
| [10] | J48, NB, SVM, BPNN | SVM | 4.1% AER |
| [11] | SVM, RF, BN | RF | 74.97% AAR |
| [12] | LR, NB, RF | RF | 0.81% EER |
| [13] | NB, KNN, RF | RF | 97.90% ACC 5.1% EER |
| [14] | BN, NB, SMOP, KNN, J48, RF | RF | 99.35% ACC |
| [15] | J48, SVM, RF, BN, NB | RF | 76% ACC |
| [16] | MLP, RF | RF | 0.01 EER |
| [17] | DTW-KNN | DTW-KNN | 5.5% Average FAR & FRR |
| [18] | SVM, RF, GB, XGB, NBB, NBG | RF, NBG | 98.35% ACC 1.88% EER |
| [19] | J48, NB, RBFN, BPNN, SVM | SVM | 3.07% AER |
| [20] | SVM | SVM | 97.1% ACC |
| [21] | SVM, RF | SVM | 79.88% ACC |
| [22] | DT, NB, RF, SVM | RF | 96.7% ACC |
| [23] | RF, XGB | RF | 1.40% FAR 2.08% FRR |
| [24] | LR, NB, KNN, DT, RF, SVM | RF | 97% ACC |
| [25] | KNN | KNN | 4.90% EER |
| [26] | DT, SVM, KNN, NBG, RF, LR | NB | 97.19% ACC |

## VI. Conclusion

Throughout the review of these papers, it was concluded that the machine learning algorithm that performs the best in touch dynamic and phone movement-based user authentication schemes are random forest and support vector machine. RF was only the best performing algorithm, but it was also found to be one of the easiest and cheapest algorithms to implement. SVM was also found to be a better choice for a classifier than other algorithms since model's using SVM can often achieve high accuracies due to SVM being able to identify more complex relationships that other algorithms cannot see. While there are many other algorithms that can perform well, RF and SVM's usability and reliability puts these algorithms ahead of the rest.

## Acknowledgment

Funding for this project has been provided by the University of Wisconsin-Eau Claire's Office of Research and Special Programs Summer Research Experience Grant. This project was done using Blugold supercomputing Cluster at uwec.